# Reversible Control of Magnetic Interactions by Electric Field in a Single Phase Material.

P. J. Ryan[1], J. -W. Kim[1], T. Birol[2], P. Thompson[3], J. -H. Lee[1], X. Ke[4], P. S. Normile[5], E. Karapetrova[1], P. Schiffer[6], S. D. Brown[3], C. J. Fennie[2], D. G. Schlom[7].

Intrinsic magnetoelectric coupling describes the interaction between magnetic and electric polarization through an inherent microscopic mechanism in a single phase material. This phenomenon has the potential to control the magnetic state of a material with an electric field, an enticing prospect for device engineering. We demonstrate 'giant' magnetoelectric cross-field control in a single phase rare earth titanate film. In bulk form, $EuTiO_3$ is antiferromagnetic. However, both anti and ferromagnetic interactions coexist between different nearest neighbor europium ions. In thin epitaxial films, strain can be used to alter the relative strength of the magnetic exchange constants. Here, we not only show that moderate biaxial compression precipitates local magnetic competition, but also demonstrate that the application of an electric field at this strain state, switches the magnetic ground state. Using first principles density functional theory, we resolve the underlying microscopic mechanism resulting in the $EuTiO_3$ G-type magnetic structure and illustrate how it is responsible for the 'giant' cross-field magnetoelectric effect.

[1] X-ray Science Division, Argonne National Laboratory, Argonne, Illinois, 60439, USA. [2] School of Applied Engineering Physics, Cornell University, Ithaca, New York 14853-1501, USA. [3] University of Liverpool, Dept. of Physics, Liverpool, L69 3BX, United Kingdom & XMaS, European Synchrotron Radiation Facility, Grenoble, France. [4] Quantum Condensed Matter Division, Oak Ridge National Laboratory, Oak Ridge, TN 37831, USA. [5] Instituto Regional de Investigación Científica Aplicada (IRICA) and Departamento de Física Aplicada, Universidad de Castilla-La Mancha, 13071 Ciudad Real, Spain. [6] Department of Physics and Materials Research Institute, Pennsylvania State University, University Park, Pennsylvania 16802, USA. [7] Department of Materials Science and Engineering, Cornell University, Ithaca, New York 14853-1501, USA. Correspondence to: pryan@aps.anl.gov

The magnetoelectric (ME) effect represents the coupling between the electric and magnetic parameters in matter[1]. Multiferroic (MF) materials with coexisting ferromagnetism (FM) and ferroelectricity were thought to offer the best prospect of achieving a strong linear ME coupling coefficient due to the combination of typically higher electric permittivity and magnetic permeability, both of which combine as an upper limit to any potential coupling strength[2]. Unfortunately, ME-MF materials are rare since ferroelectric materials need to be robustly insulating while magnetic materials are typically conducting[3]. Although uncommon, such compounds have been the subject of intense research over the past decade[1,2]. The realization of this phenomenon may lead to the development of multistate logic, new memory or advanced sensor technologies[2]. To integrate such characteristics into a functional device requires strong ME coupling between the ferroic properties, enabling the manipulation of magnetic order with an electric (E) field or electric polarity with a magnetic field. The intense search for materials exhibiting such functional ferroic control has centered on a number of complex oxide systems with a pseudo-cubic perovskite structure[2-5]. Within these systems, the electronic band structure of the central B-site cation generally determines the ferroic properties. A completely empty d band is required for ferroelectricity, while partial occupation is essential for the double and superexchange (SE) magnetic interactions, typical of these materials[3,6]. One approach circumventing this obstacle has been to engineer spatially segregated two phase systems which take advantage of electro- or magnetostriction mediated through strain or proximity to generate ME coupling[7-12]. However, finding an intrinsic single-phase mechanism would evade the inherent disadvantages and complexities of multiphase environments. In single-phase systems, the d band occupation issue is typically avoided through geometric (magnetic) frustration, where the ferroic properties arise from the Dzyaloshinski-Moriya interaction[13-16]. In these cases, the ferroic properties are weak, relegating device application unlikely.

Alternatively, the ferroic functionalities may be carried separately by the perovskite A and B site cations as in the case of BiFeO$_3$[17]. The rare earth tetravalent titanate, EuTiO$_3$ (ETO) is an emerging multi-cation ferroic prototype, whereby magnetic spins are carried by the half-occupied Eu 4f$^7$ spins and the unoccupied Ti 3d$^0$ band lends itself to potential ferroelectricity. Moreover, the anomalous response of the dielectric constant to spin alignment indicates an inherent ME coupling mechanism in ETO[18]. It was this effect that impelled Fennie and Rabe to calculate that, through strain engineering, one could create strong multiferroicity and additionally predict the exceptional strain-boundary state, allowing cross-field control capability[19]. Indeed, epitaxial films of tensile strained ETO showed multiferroicity with a large ferromagnetic moment (7μ$_B$/Eu) alongside spontaneous electric polarization (~10μCm$^{-2}$)[20]. Furthermore, ETO demonstrates a third order biquadratic ME coupling response (E$^2$H$^2$) allowing for circumvention of the linear ME susceptibility boundary condition[21].

In this article, we present E-field control of the full magnetic moment in the single phase ETO system. However, our findings do not match initial predictions. Instead, we find the dramatic ME effect does not require the proposed polar instability[19] . Rather, the combination of tuning the relative strengths of the intrinsic competing magnetic interactions under a moderate compressive strain state with the inherent paraelectric nature of the system is sufficient to generate complete ME control.

X-ray resonant magnetic scattering (XRMS) was used to confirm the magnetic structure of the contrasting strained ETO film series and reveal the emergence of competition between coexisting magnetic interactions in a moderately (-0.9%) compressed state. First principles density functional theory (DFT) calculations identified the third nearest neighbor (NN) Eu interaction central to the G-AFM structure of ETO. Finally, using *in-situ* (E-field) XRMS, we demonstrate cross-field ME control by eliminating long range AFM order and inducing a magnetic state of nanometer sized FM clusters. The underlying intrinsic mechanism is illustrated through simulations replicating the effect of E-field application by calculating the energy difference between the AFM and FM spin configurations.

**Results**

X-ray scattering is sensitive to both charge and magnetic distributions[22]. Typically, the magnetic component is about 6 orders of magnitude lower than conventional charge scattering. However, through resonance an enhanced magnetic response in the present case at the Eu $L_{II}$ edge, whereby, through the Eu 4f-5d exchange interaction, the Eu sublattice magnetic structure is probed by E1 ($2p_{1/2}$->$5d_{3/2}$) electronic excitations. Additionally, due to the polarization dependence of magnetic scattering a post sample analyzer can be used to preferentially suppress charge scattering as illustrated in Fig. 1a. Unstrained, compressive and tensile strain states were accomplished with 22nm of epitaxial cube on cube layered growth by ozone assisted molecular beam epitaxy on $SrTiO_3$(STO), $(LaAlO_3)_{0.29}$-$(SrAl_{½}Ta_{½}O_3)_{0.71}$(LSAT), and $DyScO_3$(DSO) single crystal substrates respectively[20].

Both ETO and STO share the same lattice parameter, thus when grown on the (001) surface, the film is nominally unstrained and exhibits bulk like G-AFM order with the emergence of magnetic scattering intensity at (1/2 1/2 5/2) $_{ETO}$ below $T_N$ at 5.25 K shown in Fig. 1a. A -0.9 % compressive strain is imposed by the LSAT (001) substrate and as shown in Fig. 1b also maintains G-AFM order with the onset of magnetic scattering at $T_N$ of 4.96 K. Under 1.1 % tensile strain the ETO film grown on DSO(110) is however ferromagnetic, confirmed both by the absence of a resonant magnetic signature at the (1/2 1/2 5/2)$_{ETO}$ reflection shown in Fig. 2a and with the emergence of a resonant enhancement of the magnetic scattering at the (001)$_{ETO}$ reflection at the Eu $L_{II}$ edge shown in Fig. 2b, with a $T_C$ of 4.05 K.

Contrasting with the unstrained (STO) and tensile strain (DSO) conditions the temperature dependence of the magnetic scattering intensity of the compressive state (LSAT) shows a significantly dissimilar and suppressed critical behaviour, presented in Fig. 3a. This character is found in systems due to local competition between FM and AFM interactions exemplified by the mixed-magnetic crystal system $Gd_xEu_{1-x}S$ [23]. The temperature dependent magnetic scattering intensity is fit to the critical behaviour $<m>^2 \sim I = I_0(1 - T/T_{Critical})^{2\beta}$, where $<m>$ is the magnetic moment, $I$ is the magnetic scattered intensity, $T$ = sample temperature, $T_{Critical}$ = magnetic transition temperature and $\beta$ = the critical order exponent. The AFM order of the ETO-STO film closely follows the 3 dimensional (D) Heisenberg model with a critical order exponent of $\beta$=0.385, while a larger exponent, 0.496, is found in the compressively strained ETO-LSAT film. The substantial magnetic suppression demonstrates significant local magnetic competition. Similar to the unstrained G-AFM state, the tensile strained ETO-DSO film in the FM phase also indicates 3-D Heisenberg behaviour where the local FM exchange dominates the AFM interactions without evidence of competing magnetic interactions.

Clearly both local AFM and FM interactions coexist within the ETO. In order to describe the underlying mechanism determining the G-AFM Eu spin structure, previous first principles DFT focused on the 1$^{st}$ and 2$^{nd}$ NN Eu ion interactions, illustrated in Fig. 3a–inset[24,25]. Without significant volume (lattice) expansion the calculations found FM order preferential. However to investigate the underlying factor leading to the G-AFM magnetic structure, the issue of symmetry needed to be addressed, in order to best know the structure at hand. This was accomplished through a combination of DFT calculations and XRD measurements. Until recently, bulk EuTiO$_3$ under zero stress boundary conditions was traditionally believed to be in high symmetry cubic Pm-3m space group. However our first principles calculations indicated that there were strong rotational instabilities as recently discussed by Rushanchinskii et al.[26].

The full ionic relaxations show that the lowest energy structure was I4/mcm or ($a^0a^0c^-$) in Glazer notation with the emergence of antiferrodistortive (AFD) oxygen octahedral rotations[27]. Energy gain due to this distortion is 30 eV/f.u., but the energy difference between this state and another metastable state, Imma, ($a^-a^-c^0$) is less than an eV/f.u.. The competition between these two possible rotation patterns becomes evident when we consider the structures under strain. Geometric relaxations were performed keeping the in-plane lattice constant fixed but relaxing the out-of plane lattice length corresponding to the films fixed biaxial strain boundary conditions. When we compare energies of different rotation patterns under these conditions, we see that the two aforementioned patterns compete. As a result the ($a^-a^-c^0$) pattern is favored under tensile strain and ($a^0a^0c^-$) is favored under compressive strain. In Fig. 3b we present XRD results refining the AFD related structure of the ETO film on LSAT. The combination of the non-zero H=L (1/2 5/2 1/2)$_{ETO}$ reflection with the absence of the H=K (1/2 1/2 5/2)$_{ETO}$ peak indicates the emergence of a pure in-plane AFD rotation finding agreement with the DFT calculations indicating I4/mcm or ($a^0a^0c^-$) symmetry[28]. The biaxial compressive strain effect generating the octahedral rotations is illustrated in the pseudo-cubic perovskite cell in Fig. 3b-inset.

Engaging the AFD revised strain dependent DFT calculations invoked considerable differences in the strain-phonon response compared to the initial calculations[19]. The competitive coupling between the polar and rotational structural instabilities leads to the calculated suppression of the predicted polar T01 phonon instability state[29]. The biaxial compression drives the AFD in-plane rotation in an attempt to maintain the Ti-O bond lengths, consequently preventing the T01 phonon from 'freezing' out of the film plane and thus providing an alternate avenue to minimise bond length changes. While the previous calculations without rotations (pm-3m) ($a^0a^0c^0$) indicated a ~-0.9% strain generating the polar instability, our current calculations including the AFD rotations require ~-2.5% compressive strain beyond what is currently achievable.

In table 1 we present the calculated results of the magnetic exchange interactions (J) for the 1$^{st}$, 2$^{nd}$ and 3$^{rd}$ NN Eu ions for the ETO ($a^0a^0c^-$) structure for both bulk (zero boundary conditions) and under -0.9% compressive strain, simulating epitaxial growth on the LSAT substrate. The exchange constants are broken down further into in-plane (xy) and out-of-plane (z), with positive and negative values indicating FM and AFM respectively. We find that both the 1$^{st}$ and 2$^{nd}$ NN Eu atoms interact in aggregate with FM order. The 3$^{rd}$ NN interaction however is AFM coupled. This diagonal exchange is most likely facilitated by the central Ti $3d^0$ band coupled to the Eu $4f^7$ spins through a 180° SE mechanism mediated by the intra-atomic hybridized 4f-5d orbitals, similar to the previously proposed 90° SE mechanism between the 1$^{st}$ NN Eu ions[24]. As a result, the G-AFM structure is dependent upon this 3$^{rd}$ NN interaction. Moreover, the strength of this SE coupling is reliant upon the Eu-Ti-Eu bond alignment, thus sufficient angular distortion could significantly alter the magnetic structure of the entire system[30].

Upon this premise, the paraelectric nature of the ETO film becomes central to the feasibility of ME control. In Fig. 4a, the cartoon illustrates how the 3$^{rd}$ NN interaction bond angle alignment is distorted by the Ti displacement from its central position under an applied E-field, reducing the efficacy of the interaction. Under biaxial compression the system is expected to have a preferential uniaxial polar anisotropy with the Ti displacement out of the film plane. Thus in order to examine the capability of ME cross-field control we measured the magnetic signature of the strained ETO-LSAT film where the competition between the magnetic interactions is prevalent and applied an E-field across the film to further alter the magnetic balance, as illustrated in the sample schematic in Figure 4b.

A series of reciprocal space scans through the G-AFM (1/2 1/2 5/2) $_{ETO}$ magnetic reflection at 1.9 K versus E-field strength is presented in Fig. 4c. The suppression of the XRMS intensity with E- field is clearly displayed and is ostensibly eliminated by $1.0 \times 10^5$ V/cm. The transition lacks hysteresis, is continuous, and reproducible. In Fig. 4d the resonant magnetic scattering amplitude at the fixed film Q position is plotted with decreasing E-field strength and on the return the data is extracted from a series of L scans through the magnetic reflection at each field point. This plot exemplifies the reversibility and demonstrates the stability of the transition with each data point separated by 30 minutes on the return. To further establish the proposed underlying ME microscopic mechanism we performed first principles DFT calculations to replicate the response of the applied E-field on the strained film. In Fig. 4d, the calculated enthalpy difference between the G-AFM and FM spin configurations is plotted against the effective polarization. The polarization is simulated by calculating the lowest frequency polar Eigen mode, and then forcibly and incrementally displacing the oxygen ions further from the face center ($a^0a^0c^-$) in conjunction with the Ti shift to maintain this frequency minimum. The resulting energy differences between both magnetic states are subsequently calculated. The system responds by energetically trending from AFM towards FM order with increasing polarization. Crucially it is the paraelectric ground state which allows for the ability to displace the Ti atom. This shift affects the relative strength of the local magnetic interactions reducing the 3$^{rd}$ NN exchange coupling. In order to reach a quantitative correlation between experiment and theory, we have estimated the critical E-field by dividing the energy required to displace the ions by the polarization. To generate a polarization field of P = 18 µC/cm$^2$, where AFM and FM states are degenerate, would require ~ 5 x 10$^5$ V/cm. This is comparable to the experimental field found to extinguish the AFM state, ~1.0 x 10$^5$ V/cm.

To explore the resulting induced magnetic state by E-field, we employed x-ray resonant interference scattering (XRIS)[31]. XRIS is sensitive to the magnetic moment aligned along one direction by either an internal FM interaction or an external magnetic field. Even in the AFM state, the magnetic moments uniformly canting towards an external magnetic field direction result in charge-magnetic interference of the scattered intensity illustrated in Fig. 5a. Here contrasting energy scans through the Eu L$_{II}$ edge at the (003)$_{ETO}$ reflection with opposing applied magnetic field directions ([110]$_{ETO}$) at 1.2T in the film plane demonstrate the interference effect. The measurements were made in horizontal scattering geometry which provides for additional charge scattering suppression, illustrated in Fig. 5b-inset. In ETO the AFM coupling is generally weak and as a result 1 T is sufficient to fully saturate the Eu moments along the magnetic field direction shown by the magnetic field dependence of the interference effect in Fig. 5b. This plot shows the degree of magnetic moment canting by the external H-field is proportional to the field strength.

Figure 5c-i presents the XRIS spectroscopic difference applying 0.1 T showing that the magnetic moments cant towards the field direction with ~10% of the full Eu moment. Once the electric field (1×10$^5$ V/cm) is applied, the interference effect is quenched, Fig5c-ii. This was a surprising result

because an enhanced XRIS effect due to long range FM order would be expected. Alternatively, if the electric field induced a true AFM-FM degenerate state, the magnetically frustrated moments would nevertheless align along the applied magnetic field direction resulting in an interference effect. Similarly, if the E-field caused a paramagnetic state, 0.1 T is sufficient to align the magnetic moments producing an interference effect due to small thermal fluctuations at this temperature, 1.9 K. Consequently, the magnetic state induced by the E-field is neither frustrated nor paramagnetic. However to adequately explain both the XRIS and AFM order suppression would require the emergence of short range ordered nanometer sized FM clustering. This model disrupts the long range spin coherence of the AFM order while the emergence of FM interactions in short range cluster formation remain insufficiently large to contribute to the charge-magnetic interference.

**Discussion**

Our findings present conclusive evidence for direct single phase cross-field ME control in a compressively strained $EuTiO_3$ film. Employing *in-situ* x-ray scattering measurements, we present reversible electric switching of magnetic order using a strong intrinsic coupling phenomenon. We have directly measured the microscopic magnetic structure of $EuTiO_3$ as G-AFM under low strain states (0.0% & 0.9%) and FM under 1.1% tensile strain. The magnetic critical parameters show -0.9% compressive strain alters the relative strengths of coexisting AFM and FM magnetic interactions bringing them into competition. First principles DFT calculations indicate that the $3^{rd}$ NN Eu ion superexchange interaction mediated through the central Ti ion, ultimately determines the G-AFM spin periodicity along the <111> direction. Moreover by calculating the energy of the simulated ETO-LSAT film, we have replicated our experimental findings by modeling the field induced polarization effect with controlled Ti displacements. As such, the energetic stability of the AFM order dissipates leading to the emergence of FM interactions. The underlying mechanism relies on bond alignment distortion suppressing the efficacy the $3^{rd}$ NN Eu-Ti-Eu interaction. This novel 'giant' ME coupling phenomenon will likely offer intriguing prospects to explore new types of ME functionality.

**Methods**

**X-ray Resonant Magnetic Scattering (XRMS).** XRMS measurements were performed on the 6 ID-B beamline at the Advanced Photon Source and the XMaS beamline at European Synchrotron Radiation Facility. The sample was mounted on the cold finger of a Joule-Thomson stage closed cycle helium displex refrigerator. The incident x-ray energy at 6-ID was tuned to the Eu $L_{II}$ edge by a liquid nitrogen cooled double crystal Si(111) monochromator source with a 3.3cm period undulator. The XMaS beamline is a bending magnet source and the energy selection performed with a water cooled double crystal Si(111) monochromator. All samples were oriented with respect to the substrate crystallographic axis. The films are epitaxial to their substrates so that the film diffraction peaks are easily found with respect to the substrate reciprocal matrix. The incident x-ray is linearly polarized perpendicular to the scattering plane (σ polarization). The resonant magnetic scattering, arising from electric dipole transitions from the 2$p$-to-5$d$ states, rotates the polarization resulting in π polarized photons (parallel to the scattering plane). A post sample pyrolytic graphite analyzer at the (0 0 6)$_{PG}$ reflection was used to select π-polarized radiation and suppress the background from charge scattering (σ polarized light).

Magnetic scattering is not frequently used to measure ferromagnetism with zero magnetic field because the magnetic reflection occurs at the same position in reciprocal space as the larger charge scattered intensity. In the rare-earth compound however, the magnetic scattering intensity can be comparable to the final charge scattering by coupling the large resonant enhancement and suppression of the charge scattering by polarization analysis[32]. It is expedient to choose the optimum reflection to maximize the magnetic to charge scattering ratio since the chemical structure factor is different from the magnetic structure factor. The (0 0 Odd)$_{ETO}$ charge reflection is about 40 times smaller than the (0 0 Even)$_{ETO}$ reflection, where the diffracted x-rays are in-phase enhancing the scattering amplitude due to the structure factor of the ETO film. The magnetic structure factor, on the other hand, is the same for both (0 0 Even)$_{ETO}$ and (0 0 Odd)$_{ETO}$ reflections. Hence, we obtained the clear resonant behaviour from the difference between the intensity of the (001)$_{ETO}$ reflection above and below T$_C$ presented in Fig. 2b.

**X-ray resonant Interference Scattering, (XRIS).** Measurement of ferromagnetic order can also be achieved from the interference between magnetic and charge scattering at the resonant edge[32]. Nominally in the XRMS process, the electric dipole resonance is dominant. The magnetic scattering (E1 transition) from the moments out of the scattering plane produce the same polarization as the charge scattering when the incoming polarization is parallel to the scattering plane ($\pi$-polarization). Thus, the charge scattering and magnetic scattering can interfere and consequently the interference effect is dependent on the magnetic moment direction. X-rays from synchrotron radiation are linearly polarized (in the plane of the synchrotron itself) so by using horizontal scattering geometry with the magnetic field applied in the vertical direction orthogonal to both the beam direction and beam polarization one may measure the interference change by alternating the H-field direction (up and down).

**First Principle Density Functional Theory.** We performed Density Functional Theory calculations with projector augmented wave potentials in the GGA+U framework, using VASP code. We used a 8 x 8 x 8 k-point grid for Brillouin zone integrals and a 500eV plane-wave energy cutoff. This cutoff has been increased to 600eV in certain parts of the calculations for greater accuracy. Geometric relaxations are done by keeping the in-plane lattice parameter 'a' fixed and relaxing the out-of-plane lattice parameter 'c'. Residual force threshold was decreased to 0.5 eV/Å where necessary in order to resolve differences between states close in energy. An external stress has been applied along the c axis in order to compensate for the overestimation of cell volume. Exchange parameters for an Ising model are fitted to total energy calculations done in a 2 x 2 x 2 perovskite supercell that consists of 40 atoms and 10 different magnetic configurations. Standard deviations of these exchange parameters are not reported since they are small and of no qualitative significance.

EuTiO$_3$ is predicted to be near a magnetic phase transition as a function of the on-site Hubbard repulsion parameter U[25]. In order to pick a best initial estimate of U, we calculate the Curie-Weiss constant and Neel temperature for bulk (under fixed stress boundary conditions) EuTiO$_3$, the results are presented in Table 2. It is not possible to reproduce the exact transition temperatures from first principles due to limitations of the simple mean field theory we used, and also because of the very small energy differences under consideration. However, if we pick a U that gives a $T_N$ : $T_C$ ratio close to experiment ($T_C$ is extracted from susceptibility measurements [33]), then we can get a good sense of the competition between FM and AFM states. We see that U = 5.7 eV, which is the value that was used in previous studies works well when oxygen rotations were not taken into account (19). However, once the rotations are taken into account and calculations are repeated in the relevant structure (I4=mcm), a U = 5.7eV overestimates the $T_N$ : $T_C$ ratio. To better fix the deficiencies in DFT, we instead use U = 6.2 eV as standard. Also, an intra-atomic exchange parameter J=1.0 eV is kept fixed.

**Acknowledgments:** Work at Argonne and use of beamline 6-ID-B at the Advanced Photon Source at Argonne was supported by the U. S. Department of Energy, Office of Science, Office of Basic Energy Sciences under Contract No. DE-AC02-06CH11357. The EPSRC-funded XMaS beamline at the ESRF is directed by M.J. Cooper, C.A. Lucas and T.P.A. Hase. P. S., X. K., J-H. L. and D.G. S were funded through PSU MRSEC, Grant DMR-0820404. T. B. and C. J. F. were supported by the DOE-BES under Grant No. DE-SCOO02334. P.J.R is grateful for fruitful discussions with Jonathon Lang, Steve May, John W. Freeland, Andreas Kreyssig and Yusuke Wakabayashi. Additional thanks to Michael Wieczorek & Chian Liu and Michael McDowell & David Gagliano for sample processing and sample environment engineering, respectively. We are grateful to O. Bikondoa, D. Wermeille, and L. Bouchenoire for their invaluable assistance and to S. Beaufoy and J. Kervin for additional XMaS support.

**References**

1. Fiebig, M. Revival of the magnetoelectric effect. *J. Phys. D* **38**, R123-R152 (2005).

2. Eerenstein, W., Mathur, N. D. & Scott, J. F. Multiferroic and magnetoelectric materials. *Nature* **442**, 759-765 (2006).

3. Hill, N. A. Why are there so few magnetic ferroelectrics? *J. Phys. Chem.* B. **104**, 6694–6709 (2000).

4. Cheong, S-W. & Mostovoy, M. Multiferroics: A magnetic twist for ferroelectricity. *Nat. Mater*. **6**, 13-20 (2007).

5. Spaldin, N. A., Cheong, S-W. & Ramesh, R. Multiferroics : Past, present, and future. *Physics Today* **63**, 38-43 (2010).

6. Cohen, R. E. Origin of ferroelectricity in perovskite oxides. *Nature* **358**, 136-138 (1992).

7. Eerenstein, W., Wiora, M., Prieto, J. L., Scott, J. F. & Mathur, N. D. Giant sharp and persistent converse magnetoelectric effects in multiferroic epitaxial heterostructures. *Nat. Mater*. **6**, 348-351 (2007).

8. Wu, S. M. *et al.* Reversible electric control of exchange bias in a multiferroic field-effect device. *Nat. Mater*. **9**, 756-761 (2010).

9. Chen, Y., Fitchorov, T., Vittoria, C. & Harris, V. G. Electrically controlled magnetization switching in a multiferroic heterostructure. *Appl. Phys. Lett.* **97**, 052502 (2010).

10. Zavaliche, F. *et al.* Electric Field-Induced Magnetization Switching in Epitaxial Columnar Nanostructures. *Nano Lett.* **5**, 1793-1796 (2005).

11. Choi, Y. J., Zhang, C. L., Lee, N. & Cheong, S-W. Cross-Control of Magnetization and Polarization by Electric and Magnetic Fields with Competing Multiferroic and Weak-Ferromagnetic Phases. *Phys. Rev. Lett.* **105**, 097201 (2010).

12. Tokunaga, Y. *et al.* Composite domain walls in a multiferroic perovskite ferrite. *Nat. Mater.* **8**, 558-562 (2009).

13. Fiebig, M., Lottermoser, Th., Frohlich, D., Goltsev, A. V. & Pisarev, R. V. Observation of coupled magnetic and electric domains. *Nature* **419**, 818-820 (2002).

14. Ueland, B. G., Lynn, J. W., Laver, M., Choi, Y. J. & Cheong, S.-W. Origin of Electric-Field-Induced Magnetization in Multiferroic $HoMnO_3$. *Phys. Rev. Lett.* **104**, 147204 (2010).

15. Hur, N. *et al.* Electric polarization reversal and memory in a multiferroic material induced by magnetic fields. *Nature* **429**, 392-395 (2004).

16. Kimura, T. *et al.* Magnetic control of ferroelectric polarization. *Nature* **426**, 55–58 (2003).

17. Wang, J. *et al.* Epitaxial $BiFeO_3$ Multiferroic Thin Film Heterostructures. *Science* **299**, 1719-1722 (2003).

18. Katsufuji, T. & Takagi, H. Coupling between magnetism and dielectric properties in quantum paraelectric $EuTiO_3$. *Phys. Rev. B* **64**, 054415 (2001).

19. Fennie, C. J. & Rabe, K. M. Magnetic and electric phase control in epitaxial $EuTiO_3$ from first principles. *Phys. Rev. Lett.* **97**, 267602 (2006).

20.	Lee, J. H. *et al.* A strong ferroelectric ferromagnet created by means of spin-lattice coupling. *Nature* **466**, 954-958 (2010).

21.	Shvartsman, V. V., Borisov, P., Kleemann, W., Kamba, S. & Katsufuji, T. Large off-diagonal magnetoelectric coupling in the quantum paraelectric antiferromagnet $EuTiO_3$. *Phys. Rev.* B **81**, 064426 (2010).

22.	Blume, M. Magnetic scattering of x-rays. *J. Appl. Phys.* **57**, 3615-3618 (1985).

23.	Hupfeld, D. *et al.* Element-specific magnetic order and competing interactions in $Gd_{0.8}Eu_{0.2}S$. *Europhys. Lett.* **49**, 92-98 (2000).

24.	Akamatsu, H. *et al.* Antiferromagnetic superexchange via 3*d* states of titanium in $EuTiO_3$ as seen from hybrid Hartree-Fock density functional calculations. *Phys. Rev.* B **83**, 214421 (2011).

25.	Ranjan, R., Nabi, H. S. & Pentcheva, R. First principles study of magnetism in divalent Eu perovskites. *J. Appl. Phys.* **105**, 053905 (2009).

26.	Rushchanskii, Konstantin Z., Spaldin, Nicola A. & Ležaić, Marjana. First-principles prediction of oxygen octahedral rotations in perovskite-structure $EuTiO_3$. *Phys. Rev.* B **85**, 104109 (2012).

27.	Glazer, A. M. The classification of tilted octahedra in perovskites. *Acta Cryst.* B **28**, 3384 (1972).

28.	May, S. J. *et al.* Quantifying octahedral rotations in strained perovskite oxide films. *Phys. Rev.* B **82**, 014110 (2010).

29.	Zhong, W. & Vanderbilt, David. Competing Structural Instabilities in Cubic Perovskites. *Phys. Rev. Lett.* **74**, 2587 (1995).

30.	Treves, D., Eibschütz, M. & Coppens, P. Dependence of superexchange interaction on $Fe^{3+}$-$O_2$-$Fe^{3+}$ linkage angle. *Phys. Lett.* **18**, 216-217 (1965).

31.	Brown, S. D. *et al.* Dipolar Excitations at the $L_{III}$ X-Ray Absorption Edges of the Heavy Rare-Earth Metals. *Phys. Rev. Lett.* **99**, 247401 (2007).

32.	Kim, J. W. *et al.* X-ray resonant magnetic scattering study of spontaneous ferrimagnetism. *Appl. Phys. Lett.* **90**, 202501 (2007).

33.	McGuire, T. R., Shafer, M. W., Joenk, R. J., Alperin, H. A. & Pickart, S. J. Magnetic Structure of $EuTiO_3$. *J. Appl. Phys.* **37**, 981-982 (1966).

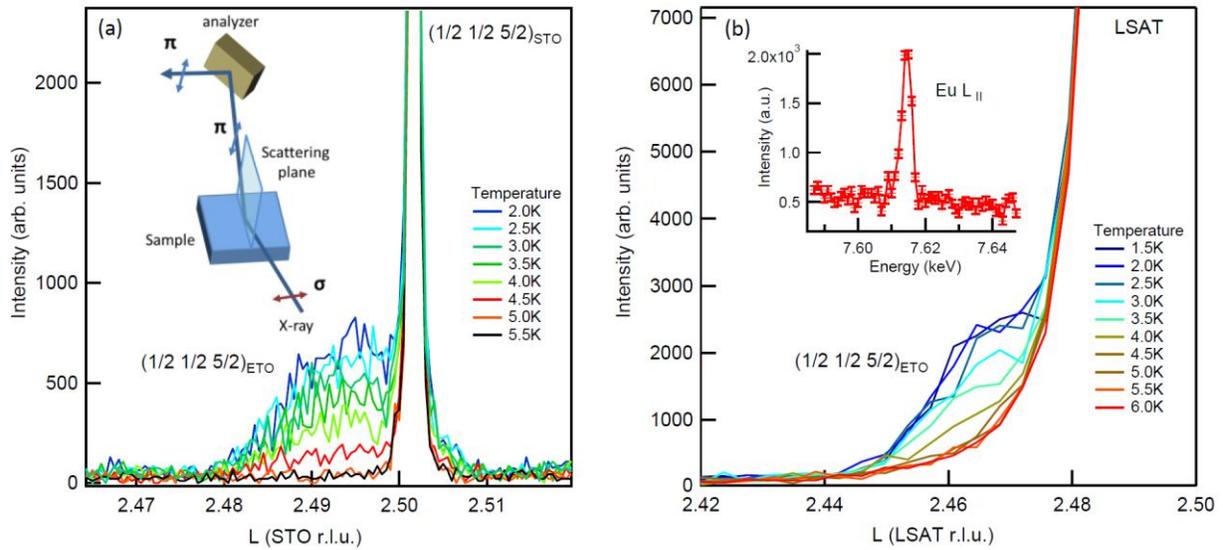

**Figure 1. X-ray resonant magnetic scattering (XRMS) presenting G-AFM order in epitaxial ETO films grown on STO(001) and LSAT (001) substrates.** (a) A series of reciprocal L-scans (film normal) through the (1/2 1/2 5/2)$_{ETO}$ reflection of the nominally unstrained ETO on STO(001) through a range of temperatures crossing $T_N$. Also seen is a half order reflection from the STO substrate due to AFD order. The measurements were taken in vertical scattering geometry with σ to π polarization selection analysis (inset) used to suppress charge and optimize the magnetic/charge scattering ratio. (b) A similar temperature dependence data set for the compressively (-0.9%) strained ETO on LSAT(001). The encumbering substrate charge intensity originates from anti-phase boundary half order reflections typical of LSAT. The inset shows the resonant response with an energy scan at 1.5K through the Eu L$_{II}$ edge.

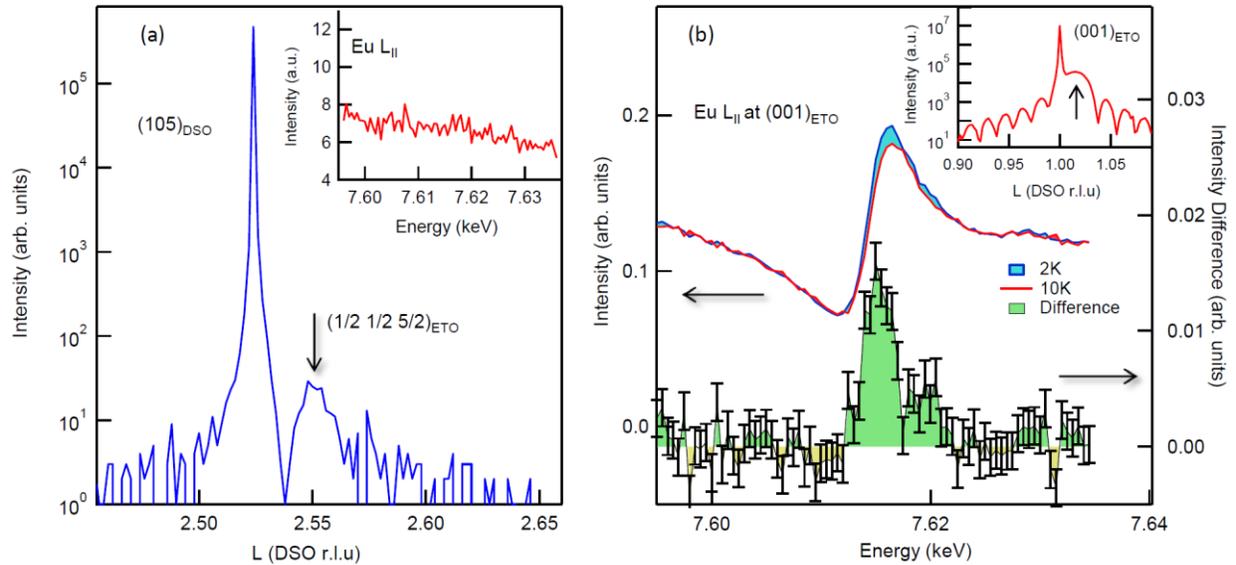

**Figure 2: XRMS demonstrating both the absence of G-AFM order and the emergence of FM order in the 1.1% tensile strained ETO on DSO(110) substrate.** a) An L-scan through the $(1/2\ 1/2\ 5/2)_{ETO}$ reflection at 1.6K below the $T_C$ mark of 4.05K. Some charge scattered leakage is detected, however an energy scan through the Eu $L_{II}$ edge is presented in the inset showing no resonant (magnetic) response. The finding demonstrates absence of long range G-AFM order of the Eu ions in the FM phase. The leaked charge amplitude derives from the octahedral tilting pattern, $(a^-a^-c^0)$ (26). (b) Presents contrasting energy scans through the Eu $L_{II}$ edge above and below $T_C$ at the integer $(001)_{ETO}$ reflection. Inset plots an L-scan through the same reflection. Due to the overlap of both charge and magnetic scattering at this reflection, suppression of the former is restricted. The onset of magnetic scattering below $T_C$ is shown with the increase of scattered intensity through the edge and indicates the spontaneous (zero field) FM long range order. The magnetic scattering contribution was about 9 % of the total intensity.

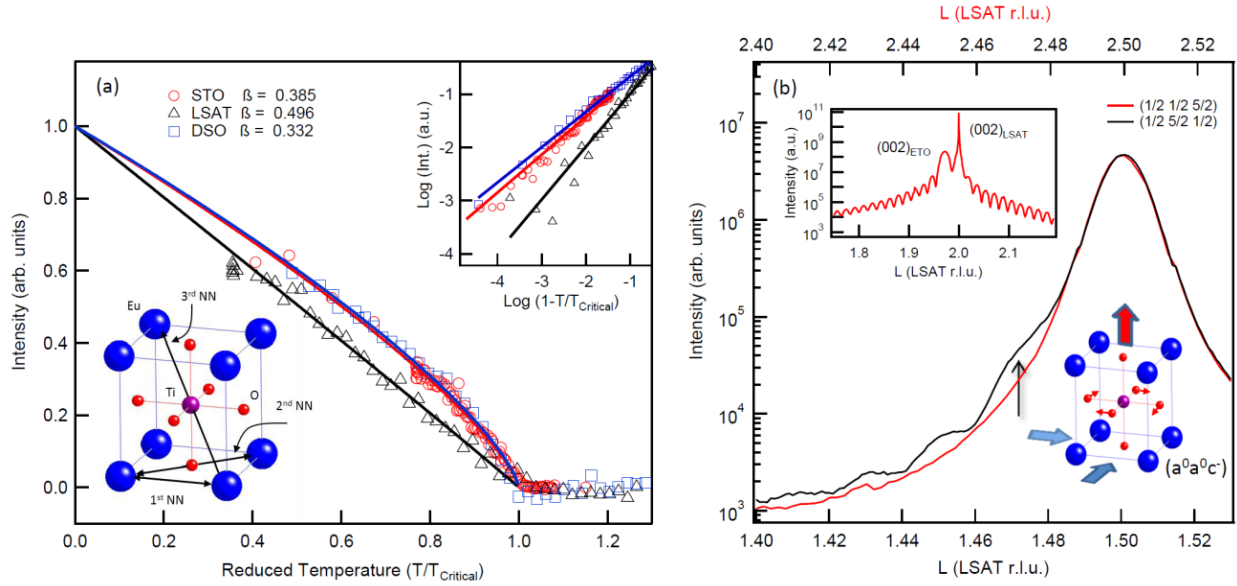

**Figure 3. Magnetic critical behaviour of the magnetic scattering intensities of the three strain states and XRD demonstrating the oxygen octahedral rotations in the ETO on LSAT.** (a) The temperature dependence of the XRMS Eu $L_{II}$ amplitudes for all three strain states, STO - unstrained, LSAT - 0.9% compressive and DSO - 1.1% tensile. The solid lines are fits of the critical behaviour $<m>^2 \sim I = I_0(1 - T/T_{Critical})^{2\beta}$, where $<m>$ is the magnetic moment, $I$ is the magnetic scattered intensity, $T$ = sample temperature, $T_{Critical}$ = magnetic transition temperature and $\beta$ = the critical order exponent. Both the G-AFM order in the unstrained (STO) and FM order of tensile (DSO) films show typical 3D Heisenberg behaviour while the compressively strained (LSAT) film shows significant suppression, a classic indicator of local magnetic competition. Inset-top, presents a log-log plot showing the near transition region. Inset bottom illustrates the multiple coexisting magnetic interactions between the 1st, 2nd and 3rd NN Eu ions. (b) The symmetry response of the ETO film to the biaxial compressive tetragonal distortion imposed by the LSAT (001) substrate. Both the (1/2 5/2 1/2)$_{ETO}$ and (1/2 1/2 5/2)$_{ETO}$ reflections at 300 K are presented. The occurrence of half order Bragg peaks show the presence of long range AFD rotations in the film. The combination of H=L allowed and H=K forbidden reflections indicate I4/mcm symmetry with the oxygen octahedral pattern ($a^0a^0c^-$) in Glazer notation[27], illustrated in the bottom inset. Again the LSAT substrate generates substantial background from the anti-phase boundary half order reflections. The top inset indicates the relative position of the (002)$_{ETO}$ reciprocal position with respect to the substrate (002)$_{LSAT}$.

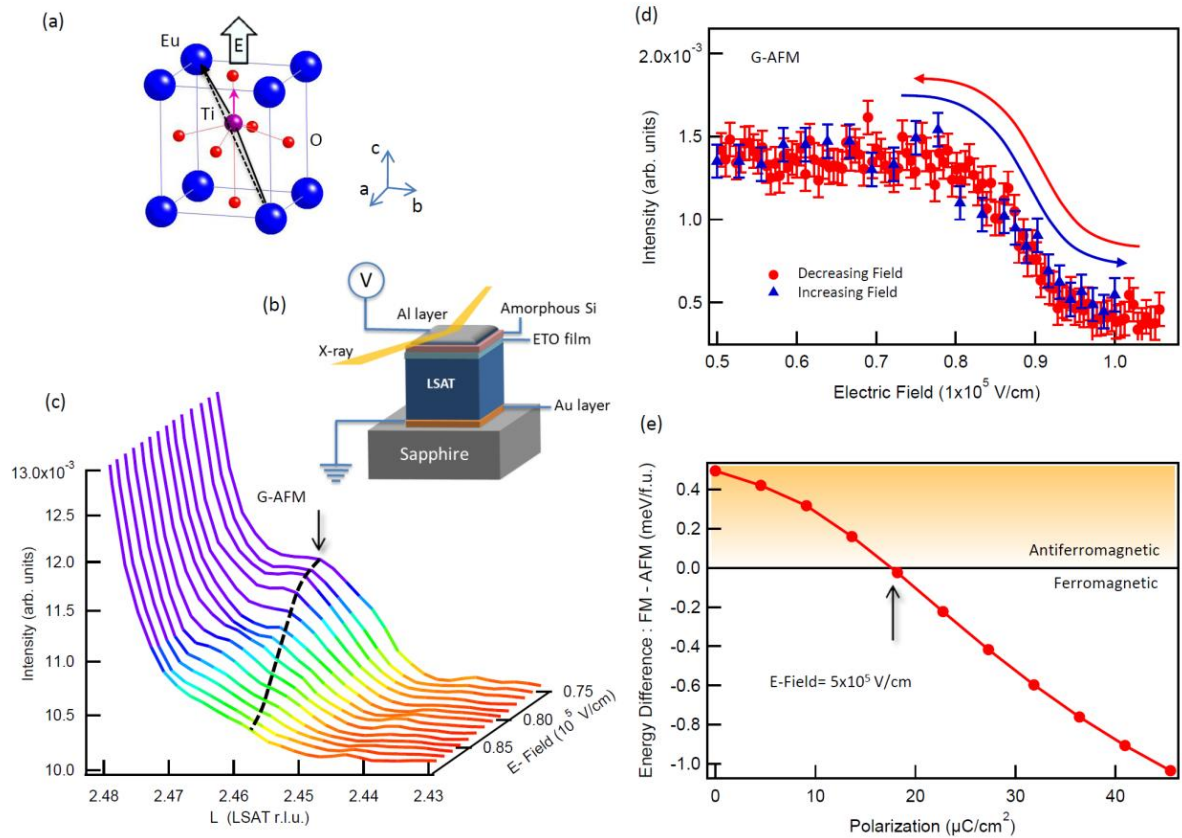

**Figure 4: With the application of an E-field across the ETO film on LSAT we demonstrate and model the control of the magnetic state.** (a) The response of the Ti atom to E-field is represented pictorially as a displacement along the direction of the field distorting the Eu-Ti-Eu 3$^{rd}$ NN bond alignment. (b) Presents a schematic of the experimental sample environment for the *in-situ* XRMS measurement. (c) A series of L-scans through the G-AFM scattered (1/2 1/2 5/2)$_{ETO}$ reflection with incrementally increasing E-field strength showing the suppressive response of the AFM signature. (d) Presents a static Q plot of the magnetic scattering intensity (1/2 1/2 5/2)$_{ETO}$ vs. E-field with increasing and decreasing field strength. (e) A plot of the first principles DFT calculations of the energy differences between FM and G-AFM spin configurations as a function of polarization modeled upon the ETO on LSAT with a compressive strain state of -0.9 %, and ($a^0a^0c^-$) octahedral symmetry[27]. The calculation replicates the suppression of the AFM state in agreement with the experimental observation.

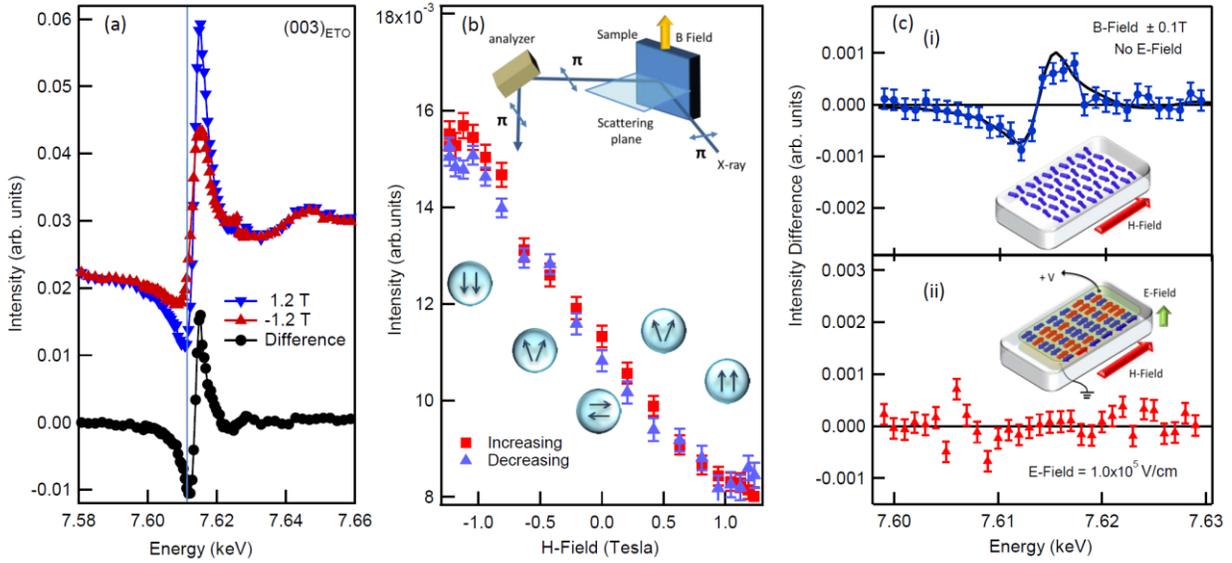

**Figure 5: The response of Eu spin alignment with applied H-Field in the ETO film on LSAT with and without E-field using X-ray Magnetic Interference Scattering (XRIS).** (a) Energy scans through the Eu $L_{II}$ edge at the $(003)_{ETO}$ reflection with ±1.2 T showing the maximum interference effect at full saturation between the magnetic and charge scattering amplitudes. The sign of the magnetic amplitude switches with the H-field direction altering the interference effect. (b) The linear XRIS-H-field dependence is presented by plotting the scattering amplitude at the line indicated energy in (a). The arrows illustrate the spin reorientation of the Eu ions with H-field and the inset shows the measurement (π-π) geometry. (c) The energy dependence of the intensity difference between ±0.1 T across the resonance edge with and without E-field application ($1 \times 10^5$ V/cm). The solid line in the top panel is a scaled version of the ±1.2 T data set. The charge-magnetic interference phenomenon is eliminated with E-field. (c-Inset) A microscopic cartoon model of the Eu spin arrangement with and without E-field. Naturally the G-AFM ordered Eu spins coherently cant towards the external H-field direction, however with applied E-field, the near magnetic degenerate states likely induce a collinear mixed AFM-FM phase devoid of long range magnetic ordering. While the FM regions produce insufficient coherency themselves, by pining

neighboring AFM spin orientations along the applied H-field direction they inhibit spin canting and in effect mute the interference effect.

**Table 1**

| ETO-LSAT | $J_{1xy}$ | $J_{1z}$ | $J_{2xy}$ | $J_{2z}$ | $J_3$ |
|---|---|---|---|---|---|
| J / $K_B$(K)-Bulk | +0.075 | -0.114 | +0.062 | +0.083 | -0.031 |
| # Neighbors | 4 | 2 | 4 | 8 | 8 |
| J / $K_B$(K)-LSAT | +0.086 | -0.147 | +0.06 | +0.087 | -0.034 |

**Table 1** lists the calculated exchange constants between the Eu ions within the unconstrained bulk I4/mcm ETO and the ETO ($a^0a^0c^-$) structure under -0.9% compressive strain, including the 1st, 2nd, and 3rd NN Eu ions describing both the in-plane (xy) and out of plane (z) interactions. Positive indicates FM and negative AFM coupling. The second row indicates the number of neighbors for each particular interaction. The 1st and 2nd NN interactions are all FM bar the 1st NN out of plane $J_{1z}$ exchange constant. The calculations indicate the importance of $J_3$ in determining the G-AFM structure in ETO.

**Table 2**

| U | 5.7eV(Pm3m) | 5.7eV(I4/mcm) | 6.0eV | 6.2eV | 6.5eV | 7.0eV | Exp. |
|---|---|---|---|---|---|---|---|
| $T_{N (K)}$ | 12.0 | 17.9 | 13.9 | 11.4 | 8.8 | 4.0 | 5.5 |
| $T_C$ | 10.5 | 5.6 | 7.0 | 7.7 | 8.3 | 9.4 | 3.8 |
| $T_N/T_C$ | 1.14 | 3.2 | 1.99 | 1.48 | 1.06 | .43 | 1.45 |

**Table 2** lists the Curie-Weiss constant, $T_C$ and $T_N$ from first principles for bulk ETO in space group I4/mcm. The last column extracts the $T_C$ from the positive magnetic susceptibility parameter in reference [33] in order to use the ratio to estimate a best guess of an appropriate value of U.